\documentclass{article}
\usepackage{spconf,amsmath,graphicx}
\usepackage{color}
\usepackage{multirow}
\usepackage{tabularx}
\usepackage[colorinlistoftodos]{todonotes}
\usepackage{url}
\usepackage{makecell}

\title{GCI detection from raw speech using a Fully-Convolutional Network}
\name{Luc Ardaillon, Axel Roebel\thanks{This work has been funded partly by the ANR projects theVoice (ANR-17-CE23-0025) and ARS (ANR-19-CE38-0001-01)}}%\thanks{Thanks to XYZ agency for funding.}}
\address{UMR 9912 STMS (IRCAM / Sorbonne University / CNRS), Paris, France}
\begin{document}
	\ninept
	\maketitle
	\begin{abstract}
		%%%%%%%%%%%%%%%%%%%% Problem formulation, definition, and applications : %%%%%%%%%%%%%%%%%%%%
		Glottal Closure Instants (GCI) detection consists in automatically detecting temporal locations of most significant excitation of the vocal tract %, occurring during the closure of vocal folds for every cycle of vocal fold vibration during the production of voiced speech, 
		from the speech signal. It is used in many speech analysis and processing applications, and various algorithms have been proposed for this purpose. %in the past for this purpose, using hand-crafted digital signal processing techniques and heuristics. 
		%%%%%%%%%%%%%%%%%%%% SOTA approaches and limitations : %%%%%%%%%%%%%%%%%%%%
		%New approaches using neural networks have been recently proposed outperforming previous approaches. \\
		%Until recently, \textcolor{green}{most/all} approaches used to be based on hand-crafted digital signal processing techniques and heuristics with various GCI candidate selection methods. %Although such approach have been shown to perform reasonably well, 
		%the quality of their results remain quite dependant on the characteristics of the analyzed speech signal and the targeted application \cite{Drugman2012, Babacan2013}. 
		Recently, new approaches using convolutional neural networks have emerged, %been proposed
		with encouraging results. %using hand-crafted digital signal processing techniques and heuristics.
		%%%%%%%%%%%%%%%%%%%% proposed approach : %%%%%%%%%%%%%%%%%%%%
		%More specifically, we first propose to use the recent PaN speech synthesis engine in order to generate a high-quality speech database with a reliable ground truth.\\
		Following this trend, we propose a simple approach %to GCI detection
		that performs a mapping from the speech waveform to a target signal from which the GCIs are obtained by peak-picking. However, the ground truth GCIs used for training and evaluation are usually extracted from EGG signals, which are not perfectly reliable and often not available. %that are usually used for establishing the ground truth for training and evaluation of such networks are not perfectly reliable and often not available. 
		%However, a particular drawback of such approaches is that the ground truth GCI used for training and evaluation are extracted from \textcolor{green}{noisy} EGG signals that are often not available, and are likely to contain errors. 
		To overcome this problem, we propose to train our network on high-quality synthetic speech with perfect ground truth.
		% we propose to train our network both on real signals with EGG-based ground truth and on synthetic speech with perfect ground truth that avoid the problems of relying on EGG signals
		%%%%%%%%%%%%%%%%%%%% evaluations : %%%%%%%%%%%%%%%%%%%% 
		%The method is compared with three existing GCI detection algorithms on two publicly available databases.\\
		%The learning capability is demonstrated with several experiments on standard datasets. \\
		The performances of the proposed algorithm are compared with three other state-of-the-art approaches using publicly available datasets, and the impact of using controlled synthetic or real speech signals in the training stage is investigated.
		%%%%%%%%%%%%%%%%%%%% claims : %%%%%%%%%%%%%%%%%%%% 
		%Additional experiment indicates that the model can perform better after trained with the speech data from the speakers who are the same as those in the test set.\\
		%The results compare well with the state-of- the-art algorithms while performing better in the case ... \\
		The experimental results demonstrate that the proposed method obtains similar or better results than other state-of-the-art algorithms %GCI detection methods
		and that using large synthetic datasets with many speakers offers a better generalization ability than using a smaller database of real speech and EGG signals.
		%Experimental results showed that the proposed method obtains similar or better results than other state-of-the-art GCI detection methods. \textcolor{green}{We further show that training the algorithm on many speakers using synthetic signals offers better generalization ability for the analysis of unseen speaker than training the algorithm on a smaller database with real signals of only 3 speakers.}
		
	\end{abstract}
	\vspace{-1em}
	\section{Introduction}
	\label{sec:Introduction}
	%%%%%%%%%%%%%%%%%%%% Problem formulation : %%%%%%%%%%%%%%%%%%%% 
	%- The Glottal Closure Instants (GCIs) are instants of significant excitations in the voice.
	Glottal Closure Instants (GCI) detection consists in finding the temporal locations of significant excitations of the vocal tract that occur in voiced speech during the closure of the vocal folds, %for every pitch cycle during the production of 
	directly from the speech signal.
	%%%%%%%%%%%%%%%%%%%% applications : %%%%%%%%%%%%%%%%%%%% 
	%As exposed in \cite{Drugman2014}, 
	GCI detection has many applications 
	%\textcolor{blue}{in speech analysis and processing}
	\cite{Drugman2014}, like the precise estimation of the %instantaneous 
	fundamental frequency ($f_0$) \cite{Yegnanarayana2009}, analysis of vocal disorders \cite{M2018, Deshpande2018}, formants estimation \cite{Anand2006}, or speech synthesis \cite{Drugman2012a}. 
	State-of-the-art $f_0$ estimators like \cite{Kim2018} or \cite{Ardaillon} work very well in most cases, but are not well adapted to deal with rough voices that contain jitter or sub-harmonics for which the pitch is not well-defined. %, as their is no clear periodicity of the signal.
	In such cases, using GCI detection instead of $f_0$ might improve the quality of resynthesized signals when used as a parameter in vocoders like SVLN \cite{Degottex2013} and PaN \cite[Section~3.5.2]{Huber2015a, ardaillon2017synthesis} for speech analysis/resynthesis and transformation. In this direction, we proposed in \cite[Section~6.3.3]{ardaillon2017synthesis} to extract pitch marks from rough singing voice extracts to estimate jitter and shimmer and use it to synthesize singing voices with rough characteristics. %\textcolor{green}{\cite{Ruinskiy2008, Verma2005} also proposed to use jitter and shimmer to transform the roughness in singing?? voice recordings.} 
	A good GCI detection would thus help to better synthesize and transform such types of voices that are currently not well handled.
	One possible way to extract the GCI positions in a speech signal is to use parallel Electroglottographic (EGG) recordings. Electroglottography uses dedicated hardware to record the vocal folds contact area by placing two contact electrodes on the speaker’s neck. GCIs can then be extracted using peak-picking on the derivative of the EGG signal. However, such recordings are rarely available, which raised the need for methods that can extract GCIs directly from speech signals.  %\textcolor{green}{(EGG can however still be used as reference for comparison.)}
	%- and EGG are subject to noise, ... so not perfectly reliable.
	%- The signal captured by the EGG allows the extraction of correct information about the GCI locations and the shape of the source signal around GCIs.
	%- However, it is preferable to have methods which extract GCIs directly from the speech signals, and use the GCI from EGG as reference for comparison.
	Many algorithms have been proposed for this purpose. Until recently, all approaches used to be based on hand-crafted digital signal processing techniques and heuristics. %with various GCI candidate selection methods.
	Thorough reviews of those techniques can be read in \cite{Drugman2014, Drugman2012, Babacan2013}, where authors compared their performances on a variety of speech and singing signals. Typically, such methods first compute an intermediate speech representation, such as the linear prediction residual \cite{Naylor2007a}, a zero-frequency filtered signal \cite{Murty2008}, or a mean-based signal \cite{Drugman2009}, which emphasizes the locations of glottal closure instants found at %/where GCI candidates are more easily manifested as
	local maxima, impulses or discontinuities. Then, dynamic-programming or peak-picking is used to select the GCIs among the detected candidates.
	%\todo{NEED TO SHORTEN INTRO FROM HERE ... }
	%For instance, the SEDREAMS algorithm \cite{Drugman2012} first determines intervals of GCI presence using a mean-based signal (obtained by calculating the mean of a sliding window over the speech signal), and then refines the GCI locations assuming that the event location corresponds to the strongest peak of the LP residual within the interval. Other examples of such approaches are the DYPSA \cite{Naylor2007a}, DPI \cite{Prathosh2013}, YAGA \cite{Thomas2012}, HE \cite{XX}, ZFR \cite{Murty2008}, or MMF \cite{Khanagha2014} algorithms.
	Examples of such approaches are the SEDREAMS \cite{Drugman2009}, DYPSA \cite{Naylor2007a}, DPI \cite{Prathosh2013}, YAGA \cite{Thomas2012}, %HE \cite{XX}, 
	ZFR \cite{Murty2008}, or MMF \cite{Khanagha2014} algorithms.
	Although such approaches have been shown to perform reasonably well, they rely on different processing techniques that require manual tuning of parameters (e.g. the mean $f_0$ value for SEDREAMS \cite{Drugman2009}), %and hand crafted heuristics that may need to be adapted to the target voice, 
	and the quality of their results remains quite dependant on the characteristics of the analyzed speech signal (e.g. pitch and voice quality, speech or singing voice) and the targeted application \cite{Drugman2012, Babacan2013}. %For instance, the SEDREAMS algorithm requires an estimate of the mean $f_0$ value to choose the proper window length. 
	In particular, it has been noted in \cite{Babacan2013} that all tested algorithms tend to perform better on speech than on singing. Moreover, some algorithms like SEDREAMS \cite{Drugman2009} or DYPSA \cite{Naylor2007a} %also rely on an external voiced/unvoiced detection method to filter out GCI candidates in unvoiced parts. 
	also detect some GCIs during unvoiced segments and thus rely on further algorithms to filter out GCI candidates in unvoiced parts. 
	%- Since all three algorithms (SEDREAMS, MMF and DYPSA) estimate GCIs also during unvoiced segments, authors recommend to filter the detected GCIs by the output of a separate voiced/unvoiced detector
	%- All these methods perform reasonably well albeit they depend largely on their choice of representative signals.
	%- \cite{Prathosh} : Further, since GCIs are present only during the voiced speech, most of the state-of-the-art GCI detectors rely on voiced-unvoiced classification as a necessary first step.
	%- \cite{Babacan2013} : Reports that the performances of algorithms like DYPSA and HE depends on the voice characteristics (pitch, voice quality, ...) and are thus not robust. Also report that all tested algorithms perform less good on singing than on speech.
	%- controlled by only one parameter (once the LP analysis is fixed): the window length. => needs the mean f0 value as a parameter.
	%- different processing techniques : As a consequence, these techniques may have different properties in terms of reliability, accuracy and robustness.
	
	%%%%%%%%%%%%%%%%%%%% Similar DNN/CNN-based approaches : %%%%%%%%%%%%%%%%%%%% 
	%To circumvent those limitations, 
	To overcome the limits of previous methods, new data-driven approaches have been recently proposed. In \cite{Matousek2017}, authors used extremely randomized trees (ERT) to classify peaks from the speech waveforms as being a GCI or not. Other recent studies proposed to use convolutional neural networks \cite{M2018, Yang2018, Goyal2019}.
	%More recently, a different data-driven approach has been proposed \cite{Matousek2017}, that uses extremely randomized trees (ERT) to classify peaks in speech waveforms as being a GCI or not. Several other recent studies proposed to use convolutional neural networks to overcome the limits of previous approaches \cite{Yang2018, M2018, Goyal2019}. 
	%Similarly to \cite{Matousek2017}, the authors in \cite{Yang2018} used a classification approach with a 2-stages procedure. First, a low-pass filtered signal is computed, whose negative peaks are taken as candidates for GCIs placement. Secondly, a CNN classification model determines for each peak whether it corresponds to a GCI or not. 
	Similarly to \cite{Matousek2017}, the authors in \cite{Yang2018} used a classification approach where GCI candidates are the negative peaks of a low-pass filtered signal. 
	Similarly, \cite{M2018} also employed a classification-based approach using 3 parallel CNNs operating on different signal representations (including the linear prediction residual). %, trained specifically on pathological speech data. 
	%In \cite{Goyal2019}, the authors formulated the GCI detection problem as a supervised multi-task learning problem using a CNN optimizing both a classification and regression cost, where a GCI is simultaneously detected and localized in a frame. 
	In \cite{Goyal2019}, the authors used a CNN to optimize both a classification and regression cost, where a GCI is simultaneously detected and localized in a frame.
	Other recent related works used regression-based approaches with neural networks for $f_0$ \cite{Kato2018} or glottal source parameters estimation (including GCI) \cite{Prathosh}. 
	However, those approaches all rely on EGG signals for establishing the ground truth reference GCIs used for training the networks. This has two main drawbacks : 1). the EGG signals are often noisy, and the extracted ground-truth GCIs %\textcolor{green}{(relying on peak-picking with manually-tuned parameters)}
	are thus likely to contain errors; 2). EGG signals are rarely available, which makes it difficult to build large multi-speaker databases for training, and thus may limit the generalization ability of the models. %\todo{ ... TO HERE.}
	%- In \cite{M2018} : CNN, but multiple columns with several type of preprocessed signals used as input. => complicated/heavy method. GCI are manually annotated from EGG : applicable because the data are very specific and the database is quite small, but not generalizable. (delay between EGG and speech is compensed. But how?). Classification-based approach with very small input frames. VERY SMALL DATASET!!
	%- EGG recordings are rarely available, and are subject to noise, ... so not perfectly reliable. 
	%- A particular drawback of such data-based approaches is that the ground truth GCI used for training and evaluation are extracted from EGG signals that are often not available and leads to errors.
	%- \cite{M2018} : also very small database
	
	%%%%%%%%%%%%%%%%%%%% Proposed approach : %%%%%%%%%%%%%%%%%%%% 
	Building up on previous work on pitch estimation of speech signals \cite{Ardaillon}  %\textcolor{green}{and taking inspiration on recent works \cite{Kato2018, Prathosh}}, 
	we propose here to use a simple and efficient Fully-Convolutional Network (FCN) to perform a mapping between the speech waveform and a target signal from which the GCIs can be easily extracted using peak-picking. %\textcolor{green}{Compared to \cite{Goyal2019}, Our formulation of the problem allows to perform both the detection and localization of the GCI in a single step, which makes the overall procedure simpler.} %(and possibly more efficient?)
	Similarly to other methods, our model can be trained on real signals when EGG recordings are available. But in order to avoid the drawbacks of relying on EGG signals, we also propose to train our network on a database of high-quality synthetic speech signals with a perfectly reliable ground truth. %for which the ground truth GCIs are perfectly known and controlled / reliable.
	%- contrary to \cite{Yang2018}, for which, a low-pass filtered signal is computed, whose negative peaks are taken as candidates for GCI placement. Secondly, a CNN-based classification model determines for each peak whether it corresponds to a GCI or not. Here we first predict a target signal, and select use peak-picking to detect the GCIs.
	%%%%%%%%%%%%%%%%%%%%  experiments and claims of this paper : %%%%%%%%%%%%%%%%%%%% 
	The performances of the proposed algorithm have been compared with three other state-of-the-art approaches using publicly available datasets, and the impact of using controlled synthetic speech or real speech signals in the training stage is explored.
	%- experiments are performed on multiple datasets comparing with four state-of-the-art algorithms to demonstrate effectiveness of the proposed method through improved detection and localization metrics.
	%- The performances of our approach have been compared with three other state-of-the-art algorithms in several experiments, using both synthetic and real signals.
	
	%%%%%%%%%%%%%%%%%%%% Organization of paper : %%%%%%%%%%%%%%%%%%%% 
	In section \ref{sec:Proposed_approach}, we will present an overview of the proposed approach and detail the architecture of our network. Then we will present in section \ref{sec:Datasets} the datasets that have been used in our experiments. Finally, the methodology and results of our evaluations will be presented in section \ref{sec:Evaluation} .
	
	\section{Proposed approach}
	\label{sec:Proposed_approach}
	\subsection{Overview}
	\label{subsec:overview}
	%\todo{This paragraph is redundant with the end of introduction. So change it here or in the intro.}
	%In this paper, we propose to adopt a similar approach to the one described in a previous article for pitch estimation of speech signals \cite{Ardaillon}\textcolor{green}{, with a few adaptations}. This approach is using a fully-convolutional network (FCN) that takes a raw speech waveform as input to perform a regression to a simple 1-dimensional target waveform from which the GCI positions can be easily extracted by means of pick-peaking. The network is then trained to minimize the binary cross-entropy between the target and predicted values.
	In this paper, we propose to use a fully-convolutional network (FCN) that takes a raw speech waveform as input to perform a mapping towards a simple 1-dimensional target waveform from which the GCI positions can be easily extracted by means of pick-peaking. The network is trained to minimize the mean-square error between the target and predicted values. The following sections detail the target signals used for the regression, the network architecture, and the procedure for GCI extraction. A python implementation of our approach, along with pretrained models, is made available online\footnote{\url{https://github.com/ardaillon/FCN_GCI}}. 
	%The proposed approach has been implemented in keras, and a python implementation with pre-trained models are made available online \textcolor{green}{[github url as a footnote]}. \todo{move this somewhere else?}
	
	\subsection{Target signals}
	\label{subsec:target_signals}
\begin{figure*}[t]
		\centering
		\includegraphics[width=\textwidth]{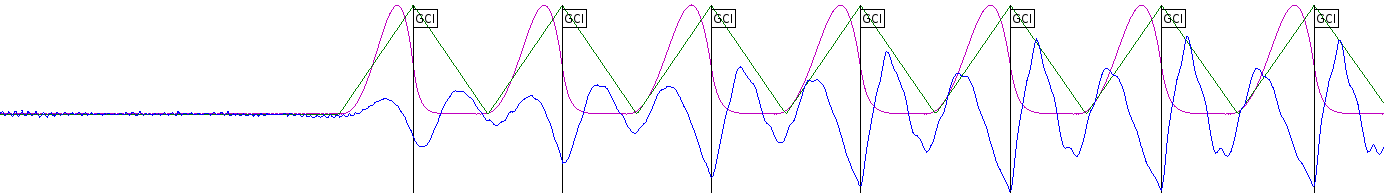}
		\caption{Example of a synthesized speech waveform (blue) with the corresponding GCI markers (black) and corresponding triangle (green) and glottal flow (purple) target signals}
		\label{fig:target_example}
	\end{figure*}

	An obvious target signal for our purpose would be a train of diracs placed at GCI positions. Initial experiments showed, however, that it is difficult for a neural network to learn such a representation, since a small variation in the input waveform (i.e. a shift by one sample) leads to a drastic change in the target. We thus need to use target signals with smoother variations. We propose to use as a target shape a triangle whose maximum of value 1 is placed at the GCI position and whose miminas of value 0 are placed at half a period to the left and right of the GCI (based on an $f_0$ analysis). %(the period being estimated based on an $f_0$ analysis of the signal).
	%simple triangle with the 0 values placed at half a pitch cycle from the GCI position
	However, besides the GCI position, this shape is not correlated to other source parameters. Using synthetic signals, a particular advantage is that the source parameters are controlled (in our case the LF model parameters \cite{Fant1985a}. See section \ref{ssec:synthDTB}). Thus, any other shape relying on those parameters might be used, the most obvious choice being the glottal flow itself. We thus propose, as an alternative to triangles, to use the glottal flow as a target shape when training our model with synthetic signals. In order to be independent of the signal's energy, each period is normalized to a maximum of 1. 
	%\textcolor{green}{Note that other target signals have been investigated, such as asymetric triangles where the 0 values are placed at the GOIs (Glottal Opening Instants), and the derivative of the glottal flow, that is typically used for synthesis, as the derivative accounts for the effect of the lips radiation. But due to space and time constraints, the results for those target won't be presented in this paper.}\todo{If we say this then we should also sumarize the result: for example, results were not significantly different from the results obtained with triangles}
	Figure \ref{fig:target_example} illustrates an example of a synthetic speech waveform with the corresponding ground truth GCIs and both the triangle and glottal flow target signals. %In this case, the target signal is a simple triangle with the 0 values placed at half a pitch cycle from the GCI position, but other shapes could be used as a target, as will be discussed in section \ref{XX}.
	
	\subsection{Network architecture}
	\label{subsec:Network_architecture}
\begin{figure*}[t]
		\centering
		\includegraphics[width=\textwidth]{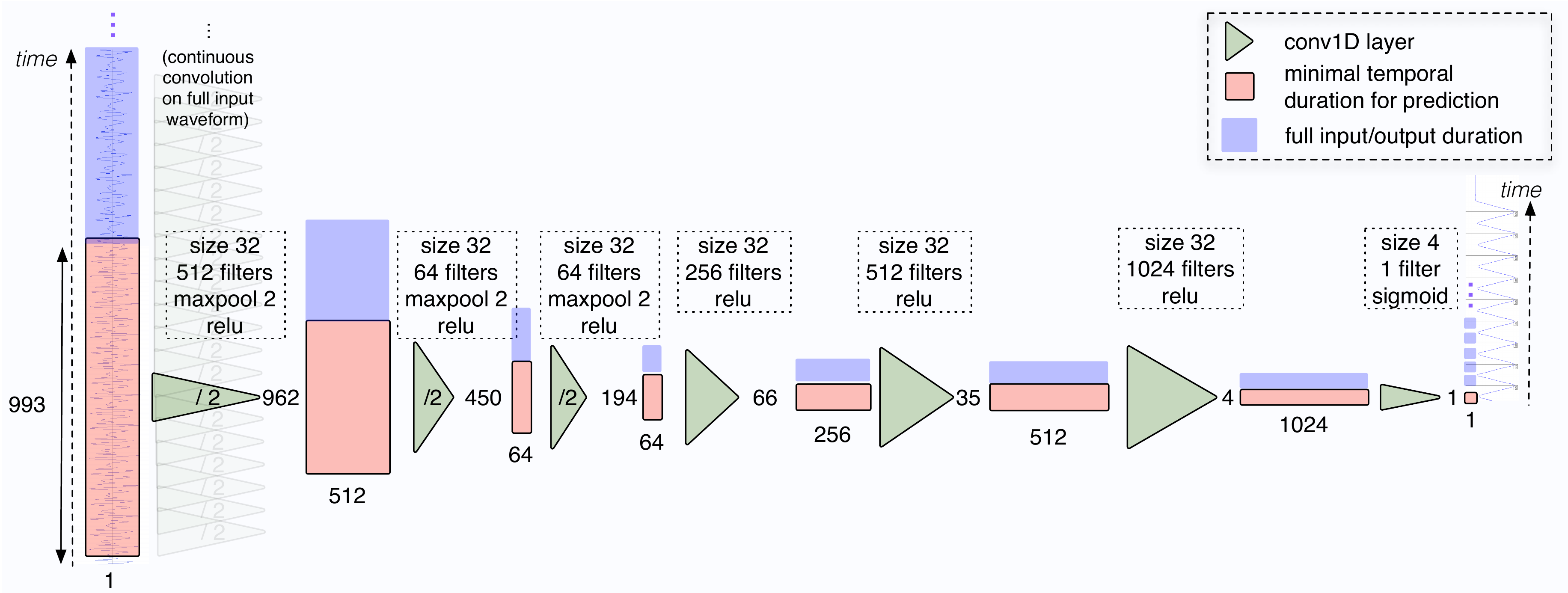}
		\caption{Architecture of the proposed fully-convolutional network. All convolutions are valid with stride 1. Minimum input size is 993.}
		\label{fig:network_archi}
	\end{figure*}

	The proposed network architecture, composed of 7 convolutional layers, is similar to the one used in \cite{Ardaillon}. Max pooling is included after the first 3 convolutional layers, each layer is followed by batch normalization, and their output are passed through relu activations, except for the last layer for which a sigmoid activation is used. The number of filters in each layer and their sizes are detailed in figure \ref{fig:network_archi}. All convolutions are valid (no padding applied) with a stride of 1. The minimal input to obtain 1 output value is a 993-samples excerpt of a speech signal sampled at 16 kHz, but any sound of size $> 993$ might be used as input. Note that the same size was used in \cite{Ardaillon}, but with a sampling rate of 8kHz.
	%993 samples is the minimum size of the input to obtain 1 value in the output vector, but any complete speech signal with a size $> 993$ samples can thus be used as input.
	%\textcolor{green}{Note that because no padding is applied before the convolutions, the temporal dimension is reduced after each convolutional layer even without max pooling. The relation between the output size lout and input size lin of a convolutional layer and the size of the filters $lfil is lout = lin - lfil + 1$. The choice of the input size of the network has thus been determined according to this relation, after the network architecture (number and positions of the convolutional and max pooling layers, and size of the filters) has been fixed.}
	Then, the only differences with \cite{Ardaillon} are the number of filters in the convolutional layers 1 to 6 that have been doubled, and the last convolutional layer that contains only 1 filter instead of 486.% (since we perform a regression instead of a classification). %\textcolor{green}{(as we want to perform a simple regression to a $1$D target vector, whereas the pitch estimation in \cite{Ardaillon} was formulated as a classification problem with 486 possible pitch classes)}.
	%\textcolor{green}{Note that the proposed fully-convolutional architecture allows to perform the inference by running the convolution in a single forward pass on the full input signal, thus saving many computations compared to frame-wise approaches that imply many redundancy in the computations.}
	
	\subsection{GCI extraction}
	\label{subsec:GCI_extraction}
	%Once the target signal has been predicted by the network, a simple peak-picking procedure is applied in order to extract the GCIs. 
	Since the network is trained on 16kHz speech signals, the predicted target at the output of the network has a sampling rate of only 2kHz, because of the down-sampling performed by the 3 max-pooling layers. %Before performing the pick-peaking, 
	For better precision, the predicted target is thus up-sampled back to 16kHz using cubic spline interpolation. Then, the GCIs are extracted using a peak-picking procedure that depends on the target signal. For the triangle target, a simple peak-picking with a minimal threshold of 0.5  %\textcolor{red}{(as is done in \cite{Goyal2019}.} 
	is used. 
	For the glottal flow target, the pick-peaking is performed on the negative peaks of its derivative,  %\todo{(with which parameters?}
	which coincide with the GCIs. %\textcolor{green}{(The detailed implementation can be found in the provided github repository).}
	%- Another difference with \cite{Ardaillon} is that the network is trained on 16kHz speech signal instead of 8kHz.
	%- in \cite{Goyal2019} : Uses a detection probability threshold of 0.5, similarly to us.
	
	\vspace{-1em}
	\section{Datasets}
	\label{sec:Datasets}
	For the purpose of proper training and objective evaluations, we need a database with a reliable ground truth. %For this purpose, the usual approach, used in \cite{XX, XX, XX} is to use
	As previously-exposed, publicly available databases with parallel speech and EGG recordings, such as the CMU artic one \cite{Kominek2004}, are typically used for this purpose \cite{Drugman2012, Prathosh2013, Yang2018}.
	%EGG measures the impedance between the vocal folds, obtained by passing a weak electrical current between a pair of electrodes placed in contact with the skin on both sides of the larynx. Based on those recordings, GCI positions can be obtained by finding the negative peaks in the derivative of the EGG signal.
	%However, the GCIs extracted from the EGG \textcolor{green}{rely on some thresholding and possibly other parameters for the peak-picking, which have to be properly tuned, and may need to be adapted for various speakers or databases which may have varying characteristics and recording conditions. Moreover, we also noticed that some of the glottal activity are often not well captured in the EGG signal. This is especially the case for the voiced consonants where the energy is low.} Finally, EGG recordings may be sometimes corrupted and are often subject to low and high-frequency noise that happen during recording, which thus require some pre-processing to first clean the signal\textcolor{green}{, as well as clics or clipping}. For those reasons, the ground truth GCI positions extracted from EGG recordings are likely to contain many errors that will impact both the training and evaluation of algorithms. 
	However, EGG recordings are often subject to low and high-frequency noise and distortions that happen during the recording. Then, depending on the quality of the signal, the pre-processing used, and the tuning of peak-picking parameters, the extracted GCIs are likely to contain many errors that will impact both the training and evaluation of algorithms. 
	
	%\textcolor{green}{Manual correction is also not applicable in this case, especially for big datasets, as the GCI position is often not easily visible in the signal.}\\
	To avoid these problems, we propose here to use re-synthesized speech signals, produced by means of a high-quality analysis/re-synthesis method achieving near transparent quality, for which the GCI positions are known and controlled (as was done in \cite{Ardaillon} for $f_0$ estimation). %as has been already proposed in \cite{Kim2018} and \cite{Ardaillon} for pitch estimation. %\todo{Don't present this as synthetic speech, but better as speech that is synthesized by means of an analysis/resynthesis method achieving near transparent speech signal quality.}
	%\textcolor{green}{Although synthetic signals may not well cover all possible cases (various voice qualities, unusual phonations, roughness, ...) that may be encountered in real speech recordings,} 
	Using such signals ensures to have a perfectly reliable ground truth exempt of any errors. Another advantage of using a synthetic database is that one can easily include many different speakers with a wide variety of signal charateristics (speech or singing, pitch, voice quality, ...), whereas it is difficult to build a database with EGG recordings for many speakers. %, since such recordings are quite rare. 
	%\textcolor{green}{Using synthetic signals might also allow including a wide variety of signal charateristics (speech and singing, with varying pitch ranges and voice qualities, ...), that might help obtaining more robust estimators across all conditions. }
	%\textcolor{green}{Another difference is that with a synthetic database, the GCI positions are perfectly aligned with the waveform. This is not the case for GCIs extracted from EGG recordings, that tend to be miss-aligned with the waveform due to the fixed (but unknown) delay time implied by the propagation of the sound wave from the vocal folds to the microphone (which would be in the order of 1.5ms for a microphone placed at a distance of 0.5m from the speaker). }
	
	\subsection{CMU artic dataset}
	\label{ssec:CMU_artic}
	Concerning the training on real speech with EGG-derived ground truth GCIs, we used the publicly-available CMU artic database \cite{Kominek2004} with the 3 speakers BDL, JMK and SLT, (similarly to \cite{Goyal2019}). 
	%Similarly to \cite{Goyal2019}, we chose to train and evaluate our model on the publicly available CMU artic database \cite{Kominek2004}, with the 3 speakers BDL, JMK and SLT. %, for which EGG recordings are provided.
	In order to obtain the ground truth GCIs, we first pre-processed the EGG signals using 5th-order high and low-pass filters with respective cutoff frequencies of 30 and 500Hz. %in order to remove the low-frequency drift and high-frequency noise.
	Then, the ground-truth GCI positions were obtained using peak-picking on the negative peaks of the derivative of the pre-processed EGG signal (as done in \cite{Babacan2013, Goyal2019}). The pick-peaking was done using the code provided by the authors of \cite{Goyal2019}, available online\footnote{\url{https://github.com/VarunSrivastavaIITD/DCNN}} (getpeaks function in dataloader.py) to have a comparable ground truth for evaluation. Then, the extracted GCI positions were used to build the corresponding triangle target signals. Note that the glottal flow target cannot be used in this case due to the missing information about the glottal pulse parameters. %\textcolor{green}{(Glottal flow is not used in this case due to missing information about the glottal source)}.%In this case, the chosen target shape is a simple triangle whose maximum is placed at the GCI position and whose miminas are placed at half a period to the left and right of the GCI (the period being estimated based on an $f_0$ analysis of the signal).
	%- (other databases for EGG are APLAWD, SAM, or PTDB-TUG, ...)
	%- 3 speakers: BDL (US male), JMK (Canadian male) and SLT (US female).
	%- SEDREAMS \cite{Drugman2} : use the CMU artic database for evaluation.
	%- In \cite{Babacan2013}, The ground truth is based on the differentiated EGG signal (dEGG). GCI marking was done on the dEGG signals by simple peak detection above an empirically-determined amplitude.
	
	%\vspace{-2em}
	\subsection{Synthetic dataset}
	\label{ssec:synthDTB}
	%As previously stated, EGG signals are rarely available, which constrains to use relatively small databases, and the extracted ground truth GCIs are likely to contain errors. We thus propose here to use synthetic signals which provide a perfectly reliable ground truth.
	%As previously stated, we also propose to use in our experiments a dataset of synthetic speech signals, in order to circumvent the drawbacks of using EGG signals. For this purpose, 
	For our synthetic dataset, we used the recent PaN vocoder described in \cite[Section~3.5.2]{Huber2015a, ardaillon2017synthesis} to resynthesize existing speech signals in a controlled way with a near-transparent quality. %PaN is a physiologically-motivated vocoder that provides high-quality synthetic speech and is well adapted in our case}. 
	Basically, the PaN vocoder uses the $f_0$ and $R_d$ parameter contours analysed on the original speech signal in order to generate a sequence of pulses based on the LF model of glottal source \cite{Fant1985a}, in an analysis/re-synthesis process. Then, the vocal tract filter is applied on the pulses %\textcolor{green}{, based on a spectral envelop estimation of the original signal} 
	and the original unvoiced component %\textcolor{green}{, extracted from the original sound,}
	is added to the voiced source to generate the complete signal. 
	In order to create this dataset, we first merged the publicly-available BREF \cite{Gauvain1990} and TIMIT \cite{Zue1990} datasets, ending up into a total of 11616 short sentences spoken by respectively 100 French and about 630 English speakers. %by many male and female voices \todo{How many of each? => Simply give the numbers to be shorter and more informative} in English and French languages. 
	Then, each sentence has been re-synthesized using the PaN synthesis engine. %\textcolor{green}{, following the procedure described in \cite{XX}}.
	In order to augment the diversity of source shapes seen by the network during the training, and thus better represent the diversity that may be encountered in real speech recordings, we also re-synthesized the same signals with $R_d$ contours shifted by fixed values of $+0.5$ and $-0.5$. %\textcolor{green}{, resulting in a total of 34848 files of synthetic speech (3 versions for each sentence and speaker)}. %\textcolor{green}{Note that the sounds have been generated at a sampling rate of 16kHz, as required by our model.} 
	During the synthesis process, we also store the GCI positions %\textcolor{green}{(set based on the $f_0$ analysis of the original signal)}
	used to produce the ground truth and the triangle target, as well as the normalized glottal flow target signals.
	%Then, the GCIs are used to generate the triangle-shaped target signal, similarly to the CMU artic database. Additionnally, we also store the glottal flow source signal that is generated during the synthesis (based on the LF model with the given $R_d$ and $f_0$ values) as an alternative target signal. 
	%As this signal is more correlated to the final speech waveform than the triangles, we expect that it may be easier for the network to predict.
	%\textcolor{green}{This is also more similar to the approach proposed in \cite{Prathosh} where the authors proposed to perform a regression from the speech waveform to the corresponding EGG signal.} Glottal flow signals also contain more information regarding other voice parameters (such as the GOI, Oq, Rd, etc...) that may also be interesting to extract, additionally to the GCIs.
	%=> regression towards glottal flow somewhat similar to \cite{Prathosh}, but more precise/reliable than using EGG as target? \cite{Prathosh} is evaluated on GCI detection.
	
	\section{Evaluation}
	\label{sec:Evaluation}
	%- The most common way to assess the performance of GCI detection techniques is to compare the estimates with the reference locations extracted from EGG signals.
	%- ERT : Use very small database to train classifier, and All speakers were part of both the training and test datasets. Reference GCIs produced by a human expert. Uses standard measures proposed in \cite{Naylor2007a} for evaluation. CMU with BDL and SLT were used for testing.
	
	\begin{table*}[]
		\label{tab:results}
		\caption{{\it Results of evaluations on various databases using standard metrics (DCNN results have been copied from \cite{Goyal2019})}}
		\footnotesize
		\centering
		\begin{tabular}{|l|l|l|l|l|l|l|l|l|l|l|l|l|}
			\hline
			\multirow{2}{*}{} & \multicolumn{3}{c|}{IDR} & \multicolumn{3}{c|}{MR} & \multicolumn{3}{c|}{FAR} & \multicolumn{3}{c|}{IDA} \\ \cline{2-13} 
			& synth          & CMU            & PTDB           & synth         & CMU           & PTDB          & synth         & CMU           & PTDB          & synth         & CMU           & PTDB          \\ \Xhline{2.5\arrayrulewidth}
			FCN-synth-tri     & 99.90          & 97.95          & 95.37          & 0.08          & 1.89          & 3.40          & \textbf{0.02} & 0.17          & 1.22          & \textbf{0.08} & 0.26          & 0.32          \\ \hline
			FCN-synth-GF      & \textbf{99.91} & 98.43          & \textbf{95.64} & \textbf{0.06} & 1.20          & 2.91          & 0.04          & 0.37          & 1.45          & 0.11          & 0.34          & 0.38          \\ \Xhline{2.5\arrayrulewidth}
			FCN-CMU-10/90     & 49.63          & 99.39          & 90.13          & 48.05         & 0.50          & 8.91          & 0.51          & 0.11          & 0.95          & 0.52          & 0.10          & \textbf{0.26} \\ \hline
			FCN-CMU-60/20/20  & 60.06          & \textbf{99.52} & 88.17          & 39.14         & 0.40          & 11.00         & 0.64          & \textbf{0.08} & \textbf{0.81} & 0.50          & \textbf{0.09} & \textbf{0.26} \\ \Xhline{2.5\arrayrulewidth}
			SEDREAMS          & 89.26          & 99.04          & 95.34          & 3.86          & \textbf{0.21} & \textbf{2.15} & 6.88          & 0.75          & 2.51          & 0.68          & 0.36          & 0.62          \\ \hline
			DPI               & 88.22          & 98.69          & 91.3           & 2.14          & 0.23          & 2.16          & 9.64          & 1.08          & 6.53          & 0.83          & 0.23          & 0.49          \\ \hline
			DCNN (from \cite{Goyal2019}) &                & 99.3           &                &               & 0.3           &               &               & 0.4           &               &               & 0.2           &               \\ \hline
		\end{tabular}
	\end{table*}
	
	\begin{table}[]
		\label{tab:results_all_GCIs}
		%\caption{{\it \textcolor{red}{Alternative evaluation considering all ground truth GCIs on CMU artic database}}}
		\caption{{\it results on CMU with modified metrics considering all GCIs }}
		\footnotesize
		\centering
		\begin{tabular}{|l|l|l|l|l|}
			\hline
			& IDR            & MR            & FAR            & IDA           \\ \hline
			FCN-synth-GF  & \textbf{92.73} & 2.86          & \textbf{10.87} & 0.47          \\ \hline
			SEDREAMS      & 90.74          & \textbf{0.19} & 61.38          & \textbf{0.30} \\ \hline
		\end{tabular}
	\end{table}
	
	\subsection{Methodology}
	\label{ssec:Methodology}
	%In \cite{Babacan2013}, the authors claimed the SEDREAMS \cite{Drugman2012} algorithm to be the more robust compared to 4 other state-of-the-art techniques, when evaluating it on various types of singing signals. In \cite{Goyal2019}, the DPI algorithm \cite{Prathosh2013} exhibits the best results behind the proposed DCNN algorithm, compared the 4 other methods. 
	%Two state of the art algorithms for GCI detection are SEDREAMS \cite{Drugman2012} and DPI \cite{Prathosh2013, Goyal2019}. 
	%We thus chose to compare the results of our approach to those 2 algorithms.
	For evaluation, we chose to compare the results of our approach to the 2 state-of-the-art algorithms SEDREAMS \cite{Drugman2012} and DPI \cite{Prathosh2013, Goyal2019}. 
	Additionally, we also use as another reference the results given in \cite{Goyal2019} for the DCNN approach. % in our evaluation. %However, due to technical and time constraints, we were not able to retrain their model on our data, and thus will only use the values given by the authors as is, trying to assess our approach in similar conditions to the ones described in their paper.\\
	To evaluate our approach in the same conditions than DCNN \cite{Goyal2019}, a first version of our model, called "FCN-CMU-10/90", has been trained on the CMU artic database using a 10/90 train/test split, where the test split is also used for validation. However, since the resulting training set is very small, we also trained a $2^{nd}$ version with more training data, called "FCN-CMU-60/20/20", using a 60/20/20 training/validation/test split. 
	%(10\%) of the CMU artic database and evaluated on the test split (90\%, also used for validation, following exactly the approach used in \cite{Goyal2019}. %as is done in \cite{Goyal2019}).
	As the glottal flow signal is not accessible for real signals, only the triangle target shape is used in this case. 
	Then, 2 additional versions of our models, called "FCN-synth-tri" and "FCN-synth-GF", have been trained on our synthetic database using respectively the triangle and the glottal flow shapes as target signals, using a 60/20/20 split in both cases. \\
	Then, all trained models, as well as the SEDREAMS and DPI algorithms, are evaluated on the tests splits of both the synthetic and CMU artic datasets. 
	%the first two models are trained on our synthetic database and will be evaluated both on the test split on the synthetic database and on the CMU artic database; 
	%\textcolor{green}{In \cite{Goyal2019} and \cite{XX?}, the models have been evaluated on speech signals from the same speakers that were used for the training.}
	Additionally, in order to assess the generalization ability of the trained models on unseen speakers, we also evaluate the algorithms on 2 speakers (M01 and F01) from the PTDB-TUG database \cite{Pirker2011}. \\
    %\textcolor{green}{Additionally, in order to assess the better generalization ability of the model trained on synthetic signal with many speakers or the model trained on real signal with only 3 speakers, we thus also evaluate the 2 model on real signals of unseen speakers from the PTDB-TUG database.}\\
    For the SEDREAMS algorithm, we used the implementation available in the COVAREP repository \footnote{\url{http://covarep.github.io/covarep/}} \cite{Degottex2014b}. The provided function requires the mean $f_0$ value as a parameter. For the CMU and PTDB databases, we thus used a value adapted to each speaker. %\textcolor{green}{(XXHz for BDL, XXHz for JMK, and XXHz for SLT). 
    However, for the synthetic database which contains many speakers, we used a single value of 150Hz. %\textcolor{red}{(or adapted to each file based on an $f_0$ analysis?)}.
    For the DPI algorithm, we used an implementation provided to us by the authors of \cite{Goyal2019}.
    %- \cite{Prathosh} : AAI method Trained on one dataset (of clean speech) and evaluated on another unseen dataset (including noise at various levels). => Obtains good and consistent performances across several dataset compared to other algorithms.

\subsection{Training procedure}
\label{ssec:Training_procedure}
In all of our experiments, the networks are trained to minimize a mean-square error loss using mini-batch gradient descent with the Adam optimizer \cite{Kingma2014}, with mini-batches composed of 128 examples randomly selected from the training set. An initial learning rate of 0.0002 is used, and a reduction by a factor 0.75 is applied when the validation loss doesn’t decrease for 10 epochs (where an epoch consists of 500 batches), with a minimum value 0.0000025. The training is stopped when the validation accuracy has not improved for 64 epochs. The input segments used for training had a fixed size of 993 samples (the target value corresponding to the middle of the segment). %\textcolor{green}{Note that while the proposed fully-convolutional architecture can accept an input sound of arbitrary length, the input segments used for training have a fixed size of 993 samples.} %\textcolor{green}{, such that the network always output a single value during the training stage}.
Contrary to \cite{Ardaillon}, no normalization is applied on the input segments, but a global normalization to a maximum level of -3dB is applied offline on each file before training and evaluation. 
%The training on the synthetic database ("FCN-synth") is done using a 60/20/20 training/validation/test split\textcolor{green}{, the evaluation being conducted on the test split}. The training on the CMU artic database is conducted using a 10/90 training/test split, where the test split is also used for validation, as is done for the \textcolor{green}{evaluation of the} DCNN model in \cite{Goyal2019}.

\subsection{Results}
\label{ssec:results}
The considered algorithms are evaluated both in terms of reliability and accuracy using standard metrics defined in \cite{Thomas2009} : Identification Rate (IDR - \% of correct detections, higher is better), Miss Rate (MR - \% of missed detections, lower is better), False Alarm Rate (FAR - \% of false insertions, lower is better) and Identification Accuracy (IDA - standard deviation of distance between the true and predicted GCIs, lower is better). Table 1 %\ref{tab:results}
summarizes the results of our evaluation for all algorithms and databases. As can be observed, at least one of our models obtains the top results in all cases, except for the MR metric. In particular, our model "FCN-CMU-10/90" obtains slightly better results than the values reported in \cite{Goyal2019} for the DCNN model for similar evaluation settings. As could be expected, the models obtain their best results on the database they have been trained on. The models trained on synthetic data ("FCN-synth") also work well on real signals, but the inverse is not true. A probable explanation for this is that the use of many speakers in the synthetic database allows to better generalize the results to unseen speakers than for the FCN-CMU models which where trained on only 3 speakers. This tends to be confirmed by the better IDR and MR results of the FCN-synth models on the PTDB database. Regarding the FCN-synth models, the use of a glottal flow target resulted in better IDR and MR values than for triangles. However, SEDREAMS and DPI still exhibit better IDR performances that our FCN-synth models on the CMU database. (Note that the better performances of the SEDREAMS algorithm here compared to the values reported in \cite{Goyal2019} might be explained by the use of a different implementation or a better tuning of the mean $f_0$ parameter).

But the metrics used in table 1 %\ref{tab:results} 
(based on code from \cite{Goyal2019} following definitions from \cite{Naylor2007a}) only consider ground truth GCIs in voiced parts of speech for which the left and right periods correspond to a range of [50-500]Hz. This thus excludes isolated pulses as well as GCIs at the extremities of voiced segments, which correspond to real glottal activity but are harder to properly detect \cite{Yang2018}. The used metrics don't account neither for the actual number of wrongly-detected GCIs that are not contained within one period around a ground-truth GCI. 
%However, evaluations using those metrics usually only consider the voiced part of the speech signals. For instance in \cite{Goyal2019}, only ground truth GCIs for wich the left and right periods correspond to a range of [50-500]Hz are considered, which excludes isolated pulses as well as GCIs at the extremities of voiced segments, which correspond to real glottal activity but are harder to properly detect \cite{Yang2018}. %This also doesn't consider the number of false detection that are not contained within 1 period around a ground truth GCI.\todo{from Axel : "I don't see why this is not covered by the above ?" => A reformuler/préciser}
These limitations thus result in better evaluation values compared to considering the ground truth as a whole, and don't fully account for the real performances of the algorithms.  %, as it doesn't reflect the capacity of the algorithm to differentiate between voiced and unvoiced segments and to properly detect small groups of isolated pulses as well as pulses found at the extremities of voiced segments \textcolor{green}{and frequencyies below 50Hz (e.g. found in creaky voice)}.
%Furthermore, this measure doesn't consider the wrongly detected GCIs that are not contained within 1 larynx (glottal) cycle around a ground truth GCI (e.g. if the algorithm detects GCIs in unvoiced parts).
As an example, we show in table 2 %\ref{tab:results_all_GCIs} 
the results of our evaluation on the CMU artic database using an alternative implementation of the metrics that considers all of the ground truth GCIs, regardless of the $f_0$ and voicing. As can be observed, this greatly impacts the results and in those conditions, our model trained on synthetic data gets better IDR and FAR measures than the SEDREAMS algorithm. The very high FAR value for SEDREAMS in this case is due to the high number of wrongly-detected GCIs output by the algorithm during unvoiced segments, that were not considered in the results of table 1. The detection of GCIs in unvoiced parts also explains the lower MR, as the algorithm doesn't miss the GCIs at extremities of voiced segments.%\ref{tab:results}.

\section{Conclusion}
\label{sec:Conclusion}
In this paper, we proposed a novel approach to GCI detection using a fully-convolutional network and a simple peak-picking procedure. In order to avoid the problems of using EGG signals, we proposed to train our model using high-quality synthetic speech. The results of our evaluations on publicly-available datasets showed that our approach performed similarly or better than other state-of-the-art algorithms in most cases, and that using a large database of synthetic signals with many speakers for training offers better generalization abilities than using a small database of real signals with EGG recordings of few speakers. In future works, more variations, like jitter of shimmer may be included into the synthetic dataset in order to better cover all possible variations encountered in real voice signals. Beside the GCI detection problem, the prediction of the glottal flow by our model might also allow to extract other useful voice source parameters.

%- opening: in future works, more variations, like jitter of shimmer may be included into the synthetic dataset in order to better cover all possible variations encountered in real voice signals. A possibility may also to perform the training both on synthetic and real signals.
%- use other target shapes (glottal flow LF model, derivative of LF model, or asymetric triangles) to extract more informations from the target signal (GOI, OQ, ...).
%- Glottal flow signals also contain more information regarding other voice parameters (such as the GOI, Oq, Rd, etc...) that may also be interesting to extract, additionally to the GCIs.
%- opening towards neural vocoder
%- Advantage of simple and efficient approach (compared e.g. to \cite{Prathosh}, that is more complicated to train).
%- Maybe including more variations like jitter or shimmer in the synthetic database might help improving the genelizability of the model learnt on those data.
%- additionnaly to detecting GCIs, the proposed algorithm can serve as a precise instantaneous f0 estimator (especially useful if sound is not really periodic, with jitter) or for rough voices), and a voiced/unvoiced segmentation algorithm at the same time.

\section{Acknowledgement}
\label{sec:Acknowledgement}
The authors would like to thank Goyal Mohit, Srivastava Varun and Prathosh AP, who authored \cite{Goyal2019}, for their valuable help in providing matlab and python codes used in this article for establishing the ground truth GCIs from the EGG recordings and for computing the metrics, as well as their implementation of the DPI algorithm.

%\vfill\pagebreak

%\section{REFERENCES}
%\label{sec:refs}

\bibliographystyle{IEEEbib}
%\bibliography{strings,refs}
\bibliography{references.bib}

\end{document}